# Cavity Enhanced Emission from Telecom Rare-Earth System in Colloidal Host

Purbita Purkayastha [1,2], Cristian Gonzalez [3], Cameron Mollazadeh [4], Chang-Min Lee [1,5], Fariba Islam [1,5], Abhijit Biswas [1,5], Michael F. Toney [4,6], Christopher B. Murray [3,7], Edo Waks [1,2,5]

[1] Institute for Research in Electronics and Applied Physics and Joint Quantum Institute, University of Maryland, College Park, Maryland 20742, USA
[2] Department of Physics, University of Maryland, College Park, Maryland 20742, USA
[3] Department of Chemistry, University of Pennsylvania, Philadelphia, Pennsylvania 19104, USA
[4] Department of Materials Science and Engineering, University of Colorado Boulder, Boulder, Colorado 80309, USA
[5] Department of Electrical and Computer Engineering, University of Maryland, College Park, Maryland 20740, USA
[6] Chemical and Biological Engineering Department, University of Colorado-Boulder, Boulder, Colorado 80309, USA
[7] Department of Materials Science and Engineering, University of Pennsylvania, Philadelphia, Pennsylvania 19104, USA

**Abstract** : Erbium incorporated in a ceria ($CeO_2$) host offers strong potential as a spin qubit platform, providing a pathway toward telecom-compatible quantum memory. But erbium is a dim emitter due to its long excited state lifetime. Methods to enhance its emission rate and efficiency through nanophotonic cavity integration are therefore essential. In this work, we demonstrate cavity-enhanced emission from erbium doped $CeO_2$. We couple colloidally synthesized erbium doped $CeO_2$ nanocrystals to a silicon nanobeam cavity and show a 30-fold improvement in emitter brightness combined with a 2-fold lifetime enhancement. We estimate a lower bound of 12 for the Purcell factor of cavity-coupled emitters after accounting for non-radiative decay. These results pave the way towards utilizing colloidally synthesized optically addressable spin qubits emitting at telecom wavelength for quantum networking and distributed quantum computing.

Erbium doped in ceria ($CeO_2$) holds significant promise for applications in quantum memories, repeaters and quantum networks.[1,2,3] Trivalent erbium ($Er^{3+}$) has an environmentally protected 4f-4f transition in the low-loss telecom C-band and an optically active spin ground state.[4,5] $CeO_2$ is a magnetically quiet host with nuclear spin free cerium and only 0.04% of natural oxygen contributes to magnetic noise. Cluster Correlation Expansion (CCE) calculations show that the isotopic combination of erbium and CeO$_2$ can exhibit spin coherence time of up to 47 ms.[6,7] This platform is also particularly promising for optoelectronic applications due to the tunable piezoelectric properties of CeO$_2$.[8]

Recent works have investigated both the optical and spin properties of Er-doped CeO$_2$.[9,10,11] $Er^{3+}$ in epitaxial CeO$_2$ showed a spin coherence time ($T_2$) of 176.4 μs at 77 mK under dynamical decoupling, indicating that spectral diffusion is the dominant source of decoherence.[12] The spin relaxation time ($T_1$) at this temperature was measured to be hundreds of milliseconds, highlighting the potential of utilizing collective electron spin relaxation as a local quantum memory. An alternate study on Er-doped colloidal CeO$_2$ demonstrated spin coherence properties with $T_2$ = 0.78 μs and $T_1$ = 1 ms at 3.6 K[13] which are equivalent to the $T_2$ and $T_1$ timescales of the epitaxial Er-CeO$_2$ system at the same temperature.[10] Colloidal nanoparticles are particularly promising because they don't require complex epitaxy and can be directly integrated on photonic structures through drop-casting.[14,15]

However, one of the main challenges of $Er^{3+}$ is its very long (millisecond) radiative lifetime due to the 4f transitions being parity forbidden.[16,17] When placed inside a crystal, the local electric field induces a partially allowed forced transition because of state hybridisation.[18] The low photon emission rate of $Er^{3+}$ necessitates an efficient collection scheme, and emission enhancement is crucial for utilizing this rare-earth ion for any practical application. Nanophotonic cavity integration has proven effective for enhancing the emission of erbium in various other hosts including CaWO$_4$ [19,20], TiO$_2$ [21–23], and YSO[24–26]. Extending this cavity approach to Er-doped CeO$_2$ could enable bright telecom emission in a host that exhibits a naturally quiet nuclear spin environment.

In this work, we report cavity-enhanced emission from colloidal erbium in CeO$_2$ nanocrystals coupled to high quality factor (Q) silicon nanobeam cavities. We utilize a tapered nanobeam cavity structure that exhibits high-Q and small mode volume, while simultaneously enabling efficient direct collection to a fiber. We couple colloidal nanocrystals to this cavity through drop casting, and observe a 30-fold enhancement in brightness and 2-fold lifetime enhancement of the coupled emitters. We also observe that annealing reduces non-radiative decay, consistent with its role in relieving dopant-induced strain and mitigating surface-charge effects. We establish a lower bound on the Purcell factor (F > 12). Our results are an important step towards utilizing this colloidal rare-earth platform as a telecom light source for spin photon entanglement in silicon.

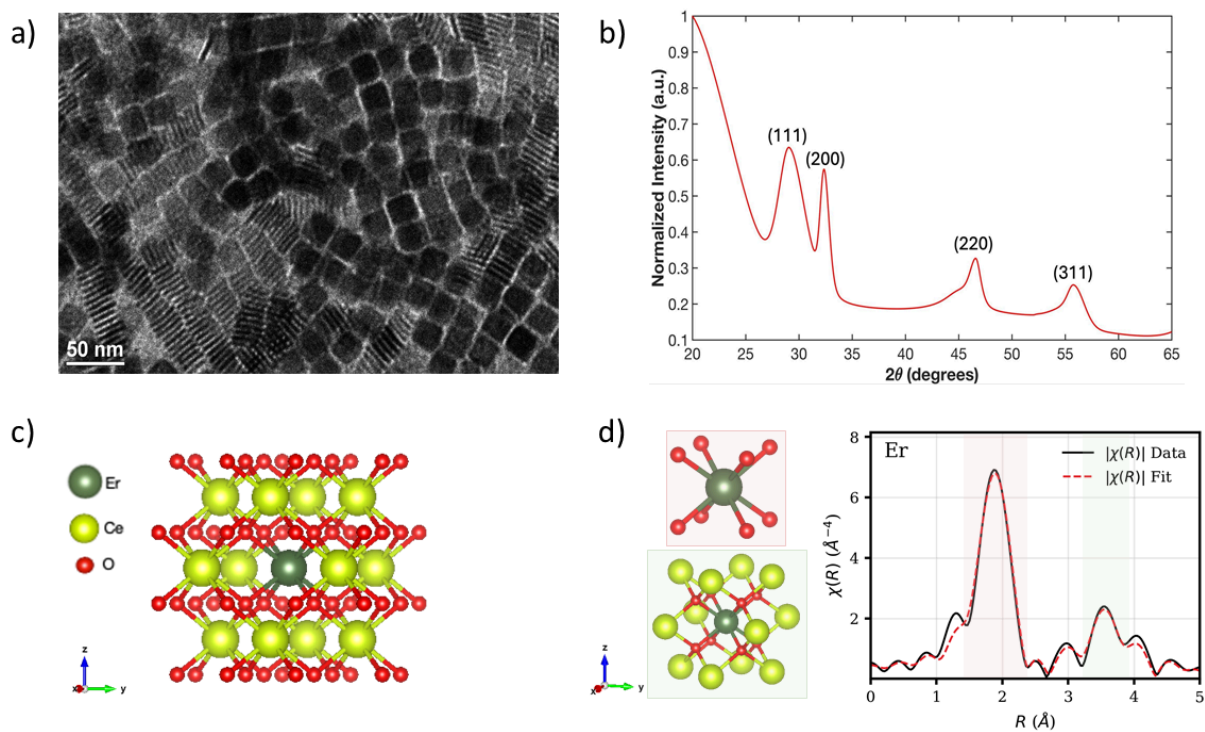

Fig1: Er-doped $CeO_2$ nanocrystals. (a) TEM image showing the anisotropic nanoplatelet morphology. (b) XRD pattern indexed to the fluorite $CeO_2$ lattice, with no secondary $Er_2O_3$ phases. (c) Structural model of substitutional $Er^{3+}$ replacing $Ce^{4+}$ sites, coordinated by oxygen atoms with nearby vacancies for charge balance. (d) EXAFS spectrum at the Er $L_3$-edge with fit, confirming 8 O neighbors in the first shell and Er–Ce interactions, consistent with substitutional doping into the host lattice.

The synthesis of Er-doped ceria nanocrystals follows the method previously reported by Wang et al.,[27] with slight modifications. The erbium precursor was co-introduced during the thermal decomposition of cerium acetate in the presence of oleic acid and oleylamine as coordinating ligands. The reaction mixture was heated to 320–330 °C under solvothermal conditions to promote the growth of doped nanoplatelets (see Supporting Information: S1 for further details).

Figure 1a shows a transmission electron microscopy (TEM) image of the resulting Er-doped ceria nanocrystals, which exhibit an anisotropic morphology characteristic of nanoplatelets. The nanocrystals display an average thickness of 4.31 ± 0.22 nm and an in-plane length of 24.7 ± 3.8 nm. The observed stacked arrangements of these particles further support their two-dimensional morphology, distinguishing them from isotropic three-dimensional nanocubes. Such stacked configurations commonly arise in drop-cast TEM preparations due to the preferential alignment of thin plates during solvent evaporation.[28,29] Figure 1b presents the X-ray diffraction (XRD) pattern, which shows distinct reflections corresponding to the (111), (200), (220), and (311) planes of the fluorite-phase $CeO_2$ lattice, confirming the formation of phase-pure ceria.[30] Notably, no additional peaks indicative of erbium oxide are observed, suggesting successful incorporation of $Er^{3+}$ ions into the ceria lattice without detectable phase segregation.

Figure 1c illustrates a structural model of the Er-doped $CeO_2$ fluorite lattice, where $Er^{3+}$ substitutes for $Ce^{4+}$ at cationic sites. In this configuration, the dopant is expected to be surrounded by eight oxygen atoms in the first coordination shell, consistent with the native Ce–O environment in fluorite ceria.[31] The incorporation of trivalent $Er^{3+}$ introduces local charge imbalance, which is typically compensated by oxygen vacancy formation in the vicinity of the dopant.[32,33] This model thus provides the structural framework for interpreting the local environment around Er as probed by EXAFS. Figure 1d shows the Fourier-transformed EXAFS spectrum collected at the Er $L_3$-edge, along with the best-fit model (black: data, blue: fit). The prominent first-shell peak at ~1.8 Å corresponds to Er–O scattering and is well fit by an average coordination number of ~8, confirming that Er occupies substitutional Ce sites in the fluorite lattice. This coordination environment is distinct from $Er_2O_3$, where Er is only 6-fold coordinated by oxygen; fitting the data to an $Er_2O_3$ reference led to unphysical parameters and significantly poorer agreement, further ruling out secondary phase segregation. A second coordination shell at ~3.6 Å corresponds to Er–Ce interactions,[34] providing additional evidence that Er is incorporated into the $CeO_2$ host rather than forming isolated oxide domains. Together, the structural model and EXAFS fitting validate the substitutional doping of Er into $CeO_2$ nanocrystals and confirm the absence of Er-rich impurity phases.(Additional details in SI, Section 2).

In order to enhance erbium emission, we create silicon nanobeam cavities with optical modes that coincide spectrally with the erbium emission band. Our cavity structure consists of a 1D air clad silicon nanobeam cavity with linearly tapered period and hole radii forming the cavity region. Figure 2a shows a schematic of the nanobeam cavity. The cavity region is highlighted in orange. We use an asymmetric one-sided cavity, where the outcoupling is controlled by the number of right holes to guide most of the light through an adiabatic taper. The taper is designed to mode match with a lensed fiber for maximum collection. The simulated optimized cavity quality factor is in the order of a million for a double sided cavity with a small cavity mode volume of $0.11 * (\lambda/n)^3$, where $\lambda$ is the resonant wavelength and n represents the refractive index. More details on the nanobeam cavity design including the farfield and mode profile can be found in Supporting Information (S3).

We fabricate the designed structures on Silicon-on-Insulator (SOI) with 220 nm of Si device layer on top of a 3μm thick $SiO_2$ layer. We use standard electron-beam lithography and fluorine-based reactive ion etching (RIE) to pattern the silicon, followed by chemical wet etching to remove the sacrificial oxide layer and suspend the structures ( described in detail in SI, section S4). We then use transfer print lithography,[35] which is a pick and place technique using Polydimethylsiloxane (PDMS) microstamp, to position the nanobeams at the edge of a carrier wafer. The nanobeam emission can be collected by a lensed fibre from the side. Finally, we dropcast the Er-doped $CeO_2$ nanocrystals carefully on the nanobeams to study the system. Figure 2b shows the SEM image of a transferred nanobeam pad at the edge of a carrier wafer. The zoomed in figure displays the cavity region with the tapered holes.

We perform all measurements in a 3.6 K cryostat with a fiber probe station for lensed fiber coupling. We excite the emitters from the top with a 1476 nm laser focused through an objective lens and collect the photoluminescence through the waveguide using a lensed fiber. We filter out the excitation laser and analyze the collected signal with a spectrometer and superconducting nanowire single-photon detectors. To measure cavity reflectivity, we use the same lensed fiber for both exciting the sample and collecting the reflected light. The Supporting Information includes a schematic and further details of the measurement setup (S5).

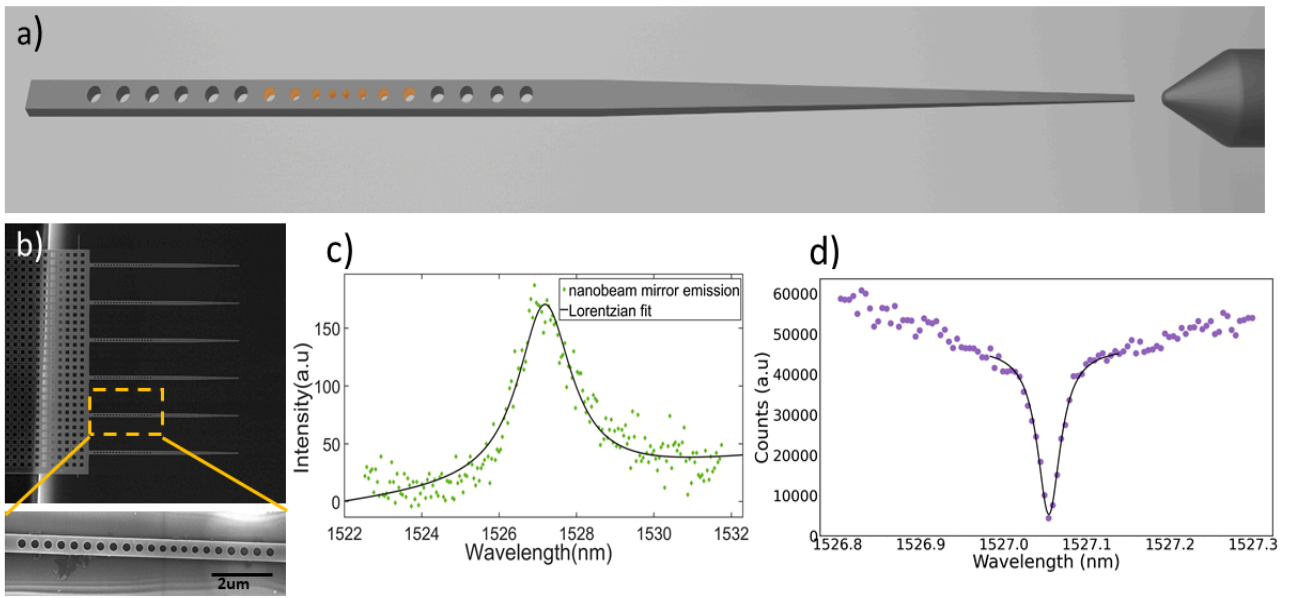

Fig 2: (a) Schematic of a nanobeam cavity design with linearly tapered cavity region marked in orange . (b) SEM image of a suspended nanobeam structure. The zoomed in region highlights the cavity structure. (c) Emission spectrum of Erbium doped $CeO_2$ nanocrystals dropcasted on a mirror nanobeam waveguide. (d) Resonant laser reflectivity spectrum of a nanobeam cavity structure for characterizing the cavity mode.

First, we perform reflectivity measurements to characterize the coupling efficiency between the nanobeams and the lensed fiber. We measure the end-to-end coupling efficiency by sending a 1500 nm laser through the lensed fiber and recording the reflected output power using a fiber coupler, for a known input excitation power (measurement setup details can be found in SI, S5(a)). We obtain a coupling efficiency ($\eta$) varying between 45% to 60% for different beams ( details in SI, S6). This variation is mostly due to slight bending of the beams based on the amount of solution drop-casted .

To study erbium's intrinsic emission properties, independent of cavity effects, we utilize a bare nanobeam waveguide structure (see SI, S3). This structure allows us to identify the emission wavelength and characteristics necessary for determining the region of interest for the cavity mode. Figure 2c displays the emission spectrum of erbium doped ceria nanocrystals dropcasted on a waveguide when excited with 100 µW of 1476 nm laser. The green points represent the measured emission and the black line shows the lorentzian fit. We observe an inhomogeneously broadened emission peak centered at 1527 nm with a FWHM linewidth of 1.7 nm.

We next characterize the nanobeam cavities by scanning a tunable laser across the cavity linewidth and measuring the reflected signal. Figure 2d shows the measured reflection spectrum, which exhibits a sharp dip at the cavity resonant wavelength of 1527.05 nm. The black line shows the lorentzian fit which gives a linewidth of 0.032 nm, corresponding to cavity quality (Q) = 46750 ± 1252. From the contrast of the reflection dip we determine that the cavity is operating near critical coupling, where out-of-plane cavity losses equal the waveguide coupling rate.

Next, we perform photoluminescence measurements on a cavity integrated to Er-doped ceria nanocrystals. We excite the cavity region with a tightly focussed 1476 nm laser. We measure the emission through the lensed fiber after filtering the scattered laser using two 1500 nm longpass filters. Figure 3a shows the measured photoluminescence spectrum. The red data points indicate the cavity enhanced emission. For comparison of brightness, we plot the emission from a waveguide, marked in green, when excited with the same pumping laser power. We normalize the emission with the coupling efficiency to the lensed fiber of the respective beams. The mirror waveguide exhibits a significantly dimmer emission and hence, the data (green points) is multiplied 10 times for better visualization. We observe a 30-fold improvement in the brightness of cavity coupled emission compared to emission from a mirror waveguide.

To characterize the radiative decay rate enhancement, we perform time-resolved photoluminescence lifetime measurements. We use an Acousto-Optic Modulator (AOM) to modulate the 1476 nm laser, generating a pulse train with a 5 ms period and 20% duty cycle. We conduct the measurements with an average top excitation power of 2 µW, well below the saturation threshold of the emitters and within the linear regime of emission intensity with respect to excitation power. We filter the emission using two 1500 nm longpass filters and a tunable fiber filter of 0.2 nm bandwidth. The signal is subsequently analysed using a single photon detector (SNSPD) and time-correlated single photon counter (PicoQuant Multiharp 160M ).

Figure 3b shows the time resolved lifetime measurement of erbium in both the cavity and bare nanobeam waveguide. The orange data points represent the decay of erbium in the cavity and the green points display the decay in the waveguide. The inset shows the pulse train used for the measurement. We fit the decays with stretched exponential function ( $I(t) = I_0 + A e^{-(t/\tau)^\beta}$ ), indicated by the solid lines in the plot. This is the most suitable fit for finding an average for heterogeneous distribution of emitters.[36] Here, I(t) indicates the photoluminescence intensity at time t, $I_0$ denotes the background, A represents the emission intensity at t=0, $\tau$ is the characteristic lifetime and β is the stretch parameter which measures the degree of inhomogeneity. The average lifetime is given by- $T_{avg} = \Gamma(1/\beta) * (\tau/\beta)$. We obtained an average lifetime ( $T_{avg}$ ) of 24.72 μs with β =0.79 for cavity coupled emitters and $T_{avg}$ = 50.58 μs with β =0.78 for the emitters in the mirror waveguide. This gives us a factor of 2 enhancement in measured lifetime due to cavity coupling.

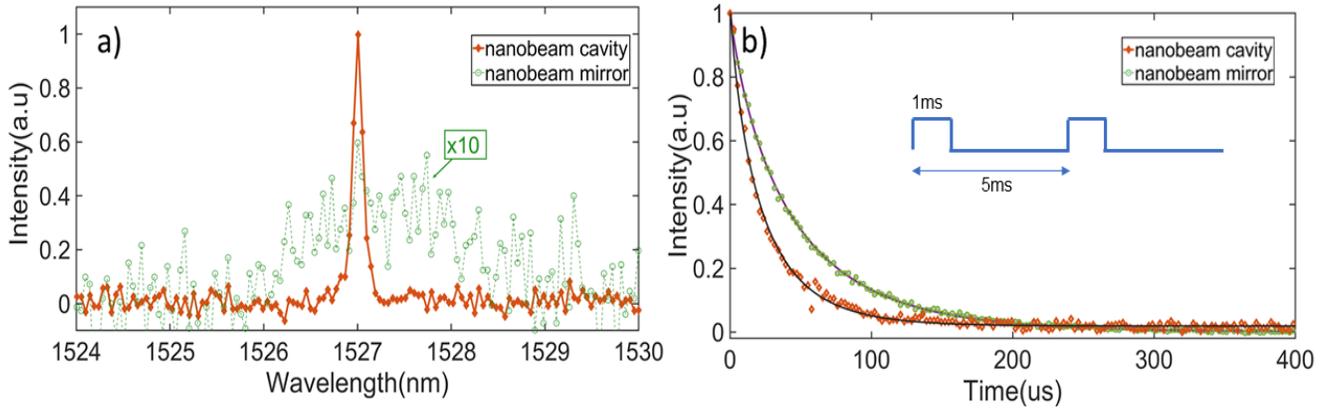

Fig3: (a) Brightness comparison of erbium-doped $CeO_2$ nanocrystals coupled to nanobeam cavity (orange points) and mirror waveguide(green points). The green data is multiplied 10 times for easier visualization. (b) Time-resolved photoluminescence decay of cavity-coupled erbium-$CeO_2$ (orange) under above band pulsed excitation and a comparison measurement for the emitters on the waveguide (green). The black lines give the stretched exponential fit of the decay curves.

To get a statistical understanding of lifetime enhancement, we compare the lifetimes in 8 different cavities and mirror waveguides. Figure 4a shows the lifetime distribution in mirror waveguides(green) and in cavities(orange). The yellow dashed line marks the center wavelength of the emission peak, while the individual cavities are slightly detuned from this center but still fall within the overall emission band. We observe that there is a clear cluster for cavity induced shorter lifetime compared to a 50 μs lifetime measured for erbium emitters in the waveguides. Figure 4b shows the lifetime of erbium in different cavities exhibiting different quality factors. We see an increasing trend of lifetime enhancement with an increase in cavity quality. For dipoles perfectly aligned with the cavity mode both spectrally and spatially, we expect a linear relationship between the quality factor and lifetime enhancement,[37] following $F \propto \frac{T_{bare}}{T_{cavity}} \propto (\frac{Q}{V})$, or equivalently $T_{cavity} \propto (\frac{1}{Q})$ where F is the Purcell factor, and $T_{bare}$ and $T_{cavity}$ are the radiative lifetimes of erbium in bulk and in the cavity, respectively. In our system, however, the dipoles are randomly oriented and exhibit spatial mismatch with the cavity mode. Also, the measured lifetime has some contribution from the non-radiative rate introduced due to the cavity. Although our measurement sample size precludes precise quantification of these effects, we still observe the expected trend.

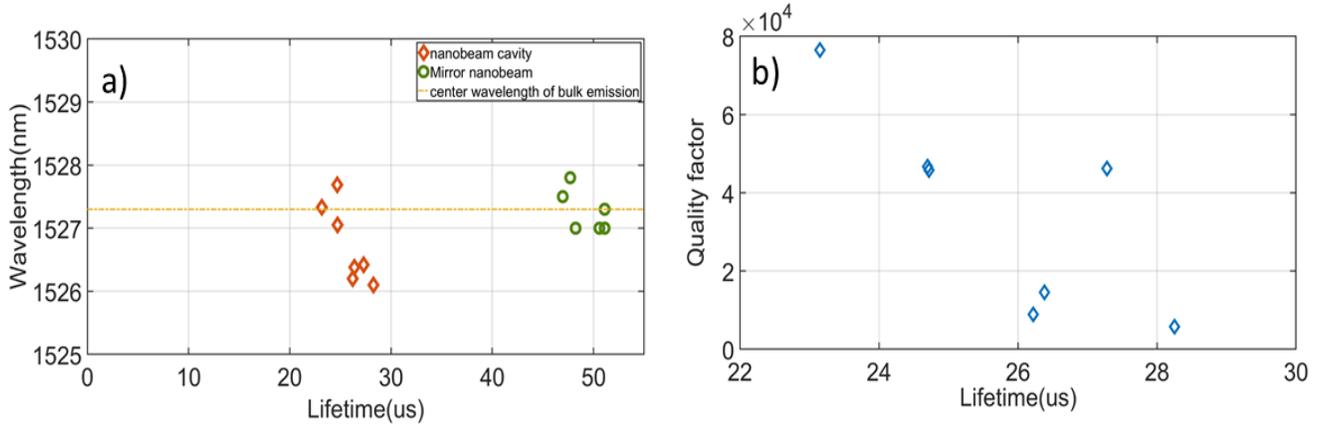

Fig 4: (a) Statistical analysis of lifetime distribution for erbium emission is mirror waveguide(green) and in nanobeam cavities (orange). The yellow dashed line displays the central wavelength of erbium emission. (b) Variation of lifetime in different cavities exhibiting different quality factors.

The observed lifetime of our emitter in waveguides (50 μs), is significantly shorter than the lifetime previously observed in CeO$_2$ (3.5 ms).[9] This reduced lifetime suggests the presence of non-radiative decay, which can arise from dangling bonds at the nanocrystal surface,[38] defects,[39] and strain-related damages while doping the nanocrystals.[40,41] To mitigate these effects, we anneal the dropcasted erbium-doped ceria nanocrystal sample at 500 degrees for 1 hr in a box furnace. We then performed time-resolved lifetime measurements on the annealed nanobeam waveguide sample (see Supporting Information, Fig:S9). The annealed sample has an average erbium emission lifetime of 320 μs with a stretch parameter β =0.51, obtained from a stretched exponential fit. This is an improvement as the measured lifetime is closer to the intrinsic erbium radiative lifetime. The high erbium density in our sample may additionally enhance non-radiative pathways through increased Er-Er interactions.[11,42] This effect can be minimized by optimizing the erbium concentration in the nanocrystals. Using the estimate of the intrinsic radiative lifetime of erbium, we can obtain a lower bound on the Purcell factor,[43] given by F > 12. (see SI, section S10)

In summary, we demonstrated enhanced emission from erbium ions doped into colloidal $CeO_2$ nanocrystals coupled to a silicon nanobeam cavity. The system exhibited a 30-fold increase in brightness and a Purcell factor exceeding 12 for cavity-coupled erbium emission. Thermal annealing was found to suppress non-radiative decay channels, underscoring the importance of post-synthesis processing. Looking ahead, several avenues can further advance this platform. Optimizing synthesis and annealing protocols will help minimize defect formation and produce highly crystalline, single-crystal $CeO_2$, thereby improving both optical and spin performance. Reducing the erbium dopant density will enable isolation of individual $Er^{3+}$ ions, a critical step toward single-emitter studies. Finally, employing resonant photoluminescence and techniques such as optically detected magnetic resonance (ODMR) will allow identification and characterization of ground-state spin-active transitions, moving closer to a spin–photon interface. Progress along these directions would constitute a major step towards establishing optically active spin qubits in a colloidal host that is easy to synthesize and integrate with photonics.

## Acknowledgements:

The authors would also like to acknowledge funding support from the National Science Foundation (grants #UWSC12985 and #ECCS2423788), and the Air Force Office of Scientific Research (grants #FA95502410266 and #FA95502210339). C.G. acknowledges support from the NSF Graduate Research Fellowship Program (NSF-GRFP),.The synthesis was supported by the NSF STC-IMOD under award DMR-2019444.

## References:

1. Awschalom, D. D., Hanson, R., Wrachtrup, J. & Zhou, B. B. Quantum technologies with optically interfaced solid-state spins. *Nat. Photonics* **12**, 516–527 (2018).

2. Saglamyurek, E. *et al.* Quantum storage of entangled telecom-wavelength photons in an erbium-doped optical fibre. *Nat. Photonics* **9**, 83–87 (2015).

3. Heshami, K. *et al.* Quantum memories: emerging applications and recent advances. *J Mod Opt* **63**, 2005–2028 (2016).

4. Dibos, A. M., Raha, M., Phenicie, C. M. & Thompson, J. D. Atomic source of single

photons in the telecom band. *Phys. Rev. Lett.* **120**, (2018).

5. Ulanowski, A., Merkel, B. & Reiserer, A. Spectral multiplexing of telecom emitters with stable transition frequency. *Sci. Adv.* **8**, eabo4538 (2022).

6. Wolfowicz, G. *et al.* Quantum guidelines for solid-state spin defects. *Nat. Rev. Mater.* **6**, 906–925 (2021).

7. Kanai, S. *et al.* Generalized scaling of spin qubit coherence in over 12,000 host materials. *Proceedings of the National Academy of Sciences* **119**, e2121808119 (2022).

8. Oh, Y. *et al.* Dielectric and piezoelectric properties of CeO2-added nonstoichiometric (Na0.5K0.5)0.97(Nb0.96Sb0.04)O3 ceramics for piezoelectric energy harvesting device applications. *IEEE Trans. Ultrason. Ferroelectr. Freq. Control* **58**, 1860–1866 (2011).

9. Grant, G. D. *et al.* Optical and microstructural characterization of Er3+ doped epitaxial cerium oxide on silicon. *APL Mater.* **12**, 021121 (2024).

10. Zhang, J. *et al.* Optical and spin coherence of Er spin qubits in epitaxial cerium dioxide on silicon. *npj Quantum Information* **10**, 1–9 (2024).

11. Inaba, T., Tawara, T., Omi, H., Yamamoto, H. & Gotoh, H. Epitaxial growth and optical properties of Er-doped CeO2 on Si(111). *Opt. Mater. Express* **8**, 2843 (2018).

12. Seth, S. K. *et al.* Spin decoherence dynamics of $Er^{3+}$ in $CeO_2$ film. *arXiv [quant-ph]* (2025).

13. Wong, J. *et al.* Coherent Erbium Spin Defects in Colloidal Nanocrystal Hosts. *ACS Nano* **18**, 19110–19123 (2024).

14. Talapin, D. V., Lee, J.-S., Kovalenko, M. V. & Shevchenko, E. V. Prospects of colloidal nanocrystals for electronic and optoelectronic applications. *Chem. Rev.* **110**, 389–458 (2010).

15. Chen, M., Lu, L., Yu, H., Li, C. & Zhao, N. Integration of colloidal quantum dots with photonic structures for optoelectronic and optical devices. *Adv. Sci. (Weinh.)* **8**, e2101560 (2021).

16. *Spectroscopic Properties of Rare Earths in Optical Materials*. (Springer, Berlin, Germany, 2010).

17. Vogler, A. & Kunkely, H. Excited state properties of lanthanide complexes: Beyond ff states. *Inorganica Chim. Acta* **359**, 4130–4138 (2006).

18. Freeman, A. J. & Watson, R. E. Theoretical investigation of some magnetic and spectroscopic properties of rare-earth ions. *Phys. Rev.* **127**, 2058–2075 (1962).

19. Ourari, S. *et al.* Indistinguishable telecom band photons from a single Er ion in the solid state. *Nature* **620**, 977–981 (2023).

20. Uysal, M. T. *et al.* Spin-photon entanglement of a single Er3+ ion in the telecom band. *Phys. Rev. X.* **15**, (2025).

21. Dibos, A. M. *et al.* Purcell Enhancement of Erbium Ions in TiO on Silicon Nanocavities. *Nano Lett* **22**, 6530–6536 (2022).

22. Ji, C. *et al.* Nanocavity-mediated Purcell enhancement of Er in TiO2 thin films grown via atomic layer deposition. *ACS Nano* **18**, 9929–9941 (2024).

23. Ji, C. *et al.* Isolation of individual Er quantum emitters in anatase TiO2 on Si photonics. *Appl. Phys. Lett.* **125**, (2024).

24. Miyazono, E., Zhong, T., Craiciu, I., Kindem, J. M. & Faraon, A. Coupling of erbium dopants to yttrium orthosilicate photonic crystal cavities for on-chip optical quantum memories. *Appl. Phys. Lett.* **108**, 011111 (2016).

25. Raha, M. *et al.* Optical quantum nondemolition measurement of a single rare earth ion

qubit. *Nat. Commun.* **11**, 1605 (2020).

26. Miyazono, E., Craiciu, I., Arbabi, A., Zhong, T. & Faraon, A. Coupling erbium dopants in yttrium orthosilicate to silicon photonic resonators and waveguides. *Opt. Express* **25**, 2863 (2017).

27. Wang, D. *et al.* Synthesis and oxygen storage capacity of two-dimensional Ceria nanocrystals. *Angew. Chem. Weinheim Bergstr. Ger.* **123**, 4470–4473 (2011).

28. Jana, S. *et al.* Stacking and colloidal stability of CdSe nanoplatelets. *Langmuir* **31**, 10532–10539 (2015).

29. Graf, R. T. *et al.* Interparticle distance variation in semiconductor nanoplatelet stacks. *Adv. Funct. Mater.* **32**, 2112621 (2022).

30. Khan, M. A. M., Khan, W., Ahamed, M. & Alhazaa, A. N. Microstructural properties and enhanced photocatalytic performance of Zn doped $CeO_2$ nanocrystals. *Sci. Rep.* **7**, 12560 (2017).

31. Elmutasim, O. *et al.* Evolution of oxygen vacancy sites in Ceria-based high-entropy oxides and their role in $N_2$ activation. *ACS Appl. Mater. Interfaces* **16**, 23038–23053 (2024).

32. Artini, C. Rare-earth-doped Ceria systems and their performance as solid electrolytes: A puzzling tangle of structural issues at the average and local scale. *Inorg. Chem.* **57**, 13047–13062 (2018).

33. Zacherle, T., Schriever, A., De Souza, R. A. & Martin, M. *Ab initio* analysis of the defect structure of ceria. *Phys. Rev. B Condens. Matter Mater. Phys.* **87**, (2013).

34. Fonda, E., Andreatta, D., Colavita, P. E. & Vlaic, G. EXAFS analysis of the $L_3$ edge of Ce in $CeO_2$: effects of multi-electron excitations and final-state mixed valence. *J. Synchrotron Radiat.* **6**, 34–42 (1999).

35. Lee, C.-M. *et al.* Bright Telecom-Wavelength Single Photons Based on a Tapered Nanobeam. *Nano Lett.* **21**, 323–329 (2021).

36. Lindsey, C. P. & Patterson, G. D. Detailed comparison of the Williams–Watts and Cole–Davidson functions. *J. Chem. Phys.* (1980).

37. Purcell, E. M. Spontaneous emission probabilities at radio frequencies. in *Confined Electrons and Photons* 839–839 (Springer US, Boston, MA, 1995).

38. Giansante, C. & Infante, I. Surface traps in colloidal quantum dots: A combined experimental and theoretical perspective. *J. Phys. Chem. Lett.* **8**, 5209–5215 (2017).

39. Inokuti, M. & Hirayama, F. Influence of energy transfer by the exchange mechanism on donor luminescence. *J. Chem. Phys.* **43**, 1978–1989 (1965).

40. Zhu, J., Liu, F., Stringfellow, G. B. & Wei, S.-H. Strain-enhanced doping in semiconductors: effects of dopant size and charge state. *Phys. Rev. Lett.* **105**, 195503 (2010).

41. Abdulwahab, K. O., Khan, M. M. & Jennings, J. R. Doped Ceria nanomaterials: Preparation, properties, and uses. *ACS Omega* **8**, 30802–30823 (2023).

42. Graf, F. R., Renn, A., Zumofen, G. & Wild, U. P. Photon-echo attenuation by dynamical processes in rare-earth-ion-doped crystals. *Phys. Rev. B Condens. Matter* **58**, 5462–5478 (1998).

43. Kim, K.-Y. *et al.* Bright Purcell-enhanced single photon emission from a silicon G center. *Nano Lett.* **25**, 4347–4352 (2025).

# Supporting Information

# Cavity Enhanced Emission from Telecom Rare-Earth System in Colloidal Host.

Purbita Purkayastha [1,2], Cristian Gonzalez [3], Cameron Mollazadeh [4], Chang-Min Lee [1,5], Fariba Islam [1,5], Abhijit Biswas [1,5], Michael F. Toney [4,6], Christopher B. Murray [3,7], Edo Waks [1,2,5]

[1]Institute for Research in Electronics and Applied Physics and Joint Quantum Institute, University of Maryland, College Park, Maryland 20742, USA
[2]Department of Physics, University of Maryland, College Park, Maryland 20742, USA
[3]Department of Chemistry, University of Pennsylvania, Philadelphia, Pennsylvania 19104, USA
[4]Department of Materials Science and Engineering, University of Colorado Boulder, Boulder, Colorado 80309, USA
[5]Department of Electrical and Computer Engineering, University of Maryland, College Park, Maryland 20740, USA
[6]Chemical and Biological Engineering Department, University of Colorado-Boulder, Boulder, Colorado 80309, USA
[7]Department of Materials Science and Engineering, University of Pennsylvania, Philadelphia, Pennsylvania 19104, USA

## Section 1: Synthesis details

**Materials:**

Cerium(III) acetate hydrate (99.99%), erbium(III) acetate hydrate (99.9%), oleic acid (90%), oleylamine (>70%), and 1-octadecene (90%) were purchased from Sigma Aldrich. Sodium pyrophosphate ($Na_4P_2O_7$) was prepared by heating sodium hydrogen phosphate ($Na_2HPO_4$, 99.8%, Fisher Chemicals) at 400 °C for 20 h to obtain the anhydrous phase. All chemicals were used as received without further purification.

**Synthesis of Er-doped ceria nanoplates:**

A slightly modified synthesis procedure from D.Wang et al[1] was carried out using standard Schlenk techniques under controlled atmospheres. 0.5 g Cerium acetate, erbium acetate (0.2 wt% relative to Ce precursor), 2.65 g sodium pyrophosphate, 5.0 mL oleic acid, 12.5 mL oleylamine, and 22.5 mL of 1-octadecene were loaded into a 100 mL three-neck round-bottom flask equipped with a thermocouple in one side neck, a cork in the other, and the central neck connected through a bump trap to a Schlenk line. The mixture was magnetically stirred and heated to 120 °C for 20 min under vacuum to remove residual moisture and allow precursor dissolution, followed by heating to 320–330 °C for 30 min with vigorous stirring to promote nanoplate growth. After cooling to room temperature with air gun then waterbath, the product was flocculated with ethanol and isolated by centrifugation. The nanoplates were washed three times with ethanol; after the first wash, excess unreacted mineralizer ($Na_4P_2O_7$) was effectively removed. The washed material was further purified by passing the dispersion through 0.200 μm PTFE filters to remove aggregates and residual mineralizer. The purified Er-doped ceria nanoplates were finally redispersed in hexane for storage and characterization. Ceria nanoplates should show a strong purple hue.

**Structural Characterization:**

Transmission electron microscopy (TEM) was performed on a JEOL-1400 instrument operated at an accelerating voltage of 120 kV. For X-ray diffraction (XRD), samples were deposited onto single- and double-side polished silicon substrates. Diffraction patterns were collected using a Rigaku SmartLab diffractometer equipped with a Cu Kα radiation source (λ = 1.5418 Å).

## Section 2 : X-ray Absorption Spectroscopy (XAS)

XAS measurements at the Er $L_3$-Edge were collected at the Beamline for Materials Measurement, 6-BM, at the National Synchrotron Light Source II (NSLS-II), Brookhaven National Laboratory. The incident X-ray beam, generated by a three-pole wiggler source, was monochromatized using a Si(111) double-crystal monochromator in pseudo channel-cut mode with step scanning. Harmonic rejection was achieved using a flat mirror with a Pt stripe, aligned to a pitch angle of 7.0 mrad relative to the beam.

The flux of the X-ray is approximately 2 x $10^{12}$ photons per second in a focused spot size of 300μm x 300μm. Spectra were collected in fluorescence mode (sample at 45° to incident beam) using a Hitachi Vortex ME7 seven-element silicon drift detector, utilizing a Xspress3 for real-time dead-time correction. Transmission spectra for a $Er_2O_3$ reference foil was measured concurrently for energy calibration. All measurements were carried out in ambient conditions, with 20 scans per Er-doped sample in order to improve signal-to-noise ratio due to the low doping concentration of Er.

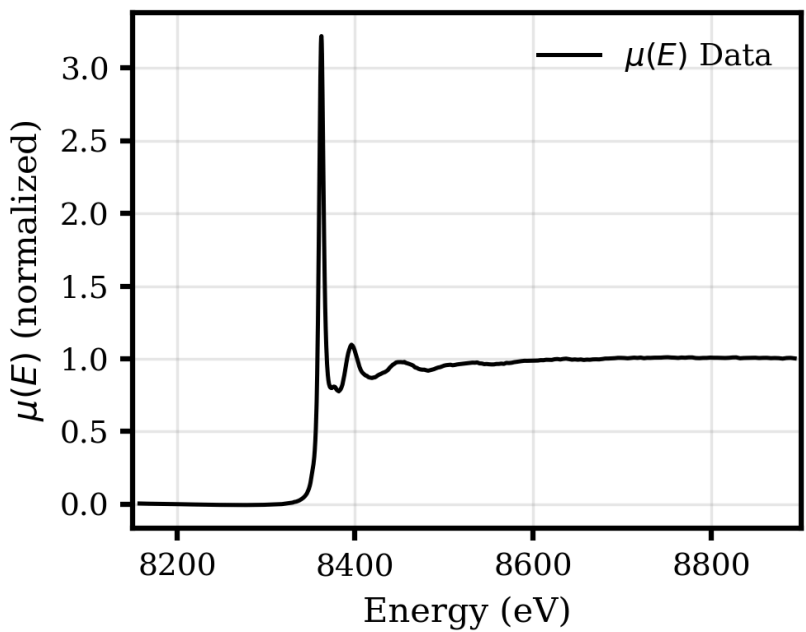

Figure S2: (a) Normalized mu(e) of Er L3 edge fluorescence data

**EXAFS Analysis :**

Raw XAS data was processed and analyzed with ATHENA and ARTEMIS from the Demeter software package.[2] Energy calibration was carried out using the maximum of the first derivative from the reference foil data to set the absorption edge energy ($E_0$). EXAFS fits were carried out using a range of 2.7 - 10.5 Å$^{-1}$ and an R range of 1.2 - 4.8 Å with $k$ weightings of 1, 2 and 3. The FEFF input used was based on the cubic fluorite structure of $CeO_2$ (space group $Fm\bar{3}m$).[3] However, to account for the Er doping into the host matrix, Method 1 as outlined in the Demeter/Artemis documentation on modeling dopants in EXAFS fitting [4] is employed. The $S_0^2$ value was determined for this fit by fitting the reference foil. To reduce the number of free parameters in the fit, $\Delta E_0$ was constrained as a single shared variable across all scattering paths. Both single-scattering (SS) and multiple-scattering (MS) contributions were included in the fit to achieve a physically realistic model. For the SS paths (Er-O and Er-Ce) $\sigma^2$, N, and $\Delta R$ were all refined. For the first MS path, only $\Delta R$ was allowed to vary; the coordination number was fixed and assumed constant, and $\sigma^2$ was fixed at twice the value of the first shell Er-O $\sigma^2$. For the second MS path, both $\sigma^2$ and $\Delta R$ were refined, as no reliable physical constraint could be applied to $\sigma^2$. This was due to one of the oxygen atoms in the path occupying a crystallographically distinct environment from that of the first shell Er-O bond.[5] The coordination number for this path was held fixed. Table 1 shows the EXAFS fit parameters.

| Complex | Scattering Path | N | $\sigma^2$ ($10^{-3}$ Å$^2$) | R (Å) | $\Delta E_o$ (eV), k-range (Å$^{-1}$), R-range (Å), R-factor Reduced $\chi^2$ |
|---|---|---|---|---|---|
| Er:CeO$_2$ | Er-O | 8.0 ± 0.6 | 9.1 ± 1.3 | 2.34 ± 0.01 | 1.2 ± 0.7, |
|  | Er-O-O | 24 | 18.2 ± 1.3 | 3.75 ± 0.17 | 2.7 - 10.5, |
|  | Er-Ce | 9.0 ± 4.0 | 14.3 ± 4.3 | 3.83 ± 0.01 | 1.5 - 4.8, |
|  | Er-O-O | 48 | 1.0 ± 0.2 | 4.90 ± 0.04 | 0.006, 54.4 |

Table 1 : EXAFS fit parameters

**EXAFS Re[χ(R)] + |χ(R)| fits**

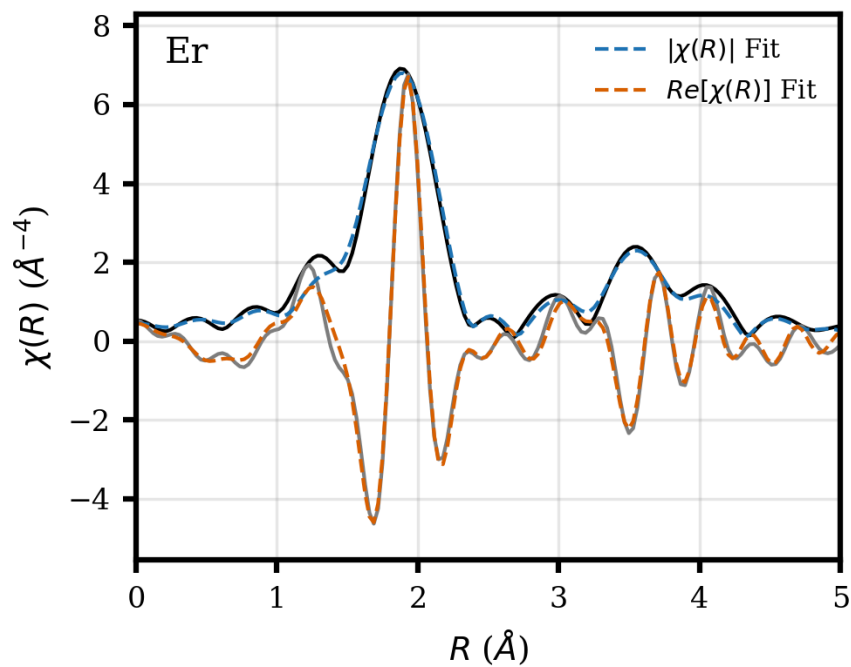

Figure S2: (b) Overlay of |χ(R)| and Re[χ(R)] data (black and grey respectively) with their corresponding fits, blue dotted and orange dotted line respectively.

**EXAFS x(k) fits**

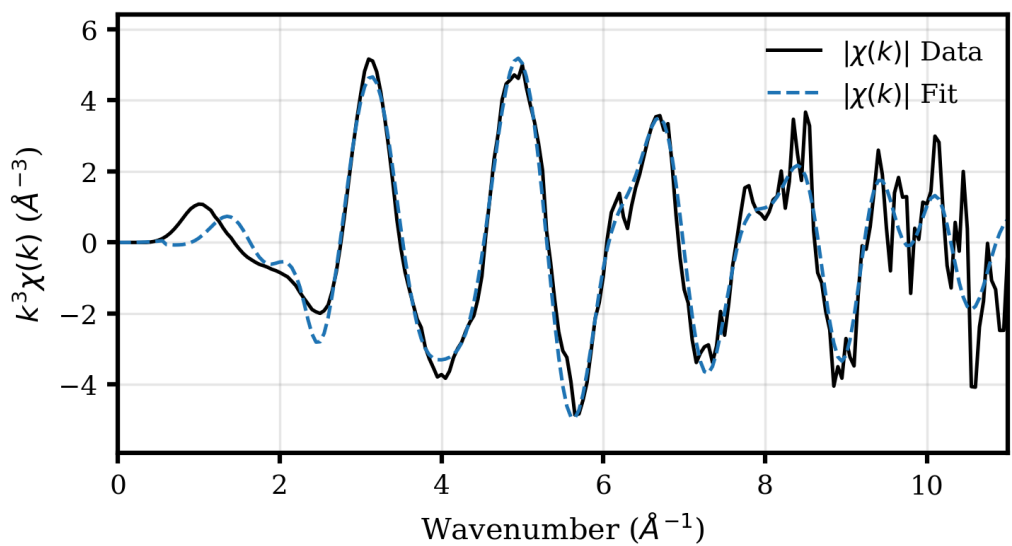

Figure S2: (c) χ(k) data (black) and χ(k) fit (blue) superimposed.

# Section 3: Nanobeam cavity design and Finite-Difference Time-Domain (FDTD) simulation of cavity mode

We use a Maxwell's equation solver (FDTD) to design our one dimensional air clad Si nanobeam photonic crystal cavities in order to realize a mode with a high quality factor at the erbium resonance. We design our structures in Si, which has a large bandgap at telecom and minimal absorption loss in Er emission wavelength. Our design consists of an array of air holes with the cavity region formed by linear tapering of both hole radii and periodicity of the central 4 holes. The in plane coupling on both sides is determined by the number of holes on left ($N_L$) and on right ($N_R$). The mirror strength is determined by $N_L$. The cavity Q is found to be nearly the same for N >= 13. In our design, we have used $N_L$ = 13 and chosen $N_R$ = 8 for reaching critical coupling condition where outcoupling through the one sided mirror is equal to scattering losses. The mode is guided through the adiabatic taper on right with $L_{taper}$ = 14μm that can be coupled to a lensed fiber. From the simulations, we obtain an optimum value of lattice constant a= 420nm, and a hole radius r = 126 nm, with the smallest hole radius in the cavity region $r_4$ = 102 nm, beam width b=520 nm. Fig S3b shows the electric field distribution of the cavity mode. Fig S3c depicts the far-field profile of the output field transmitted through the taper, exhibiting a Gaussian distribution that is mode-matched to the lensed fiber to maximize collection efficiency.

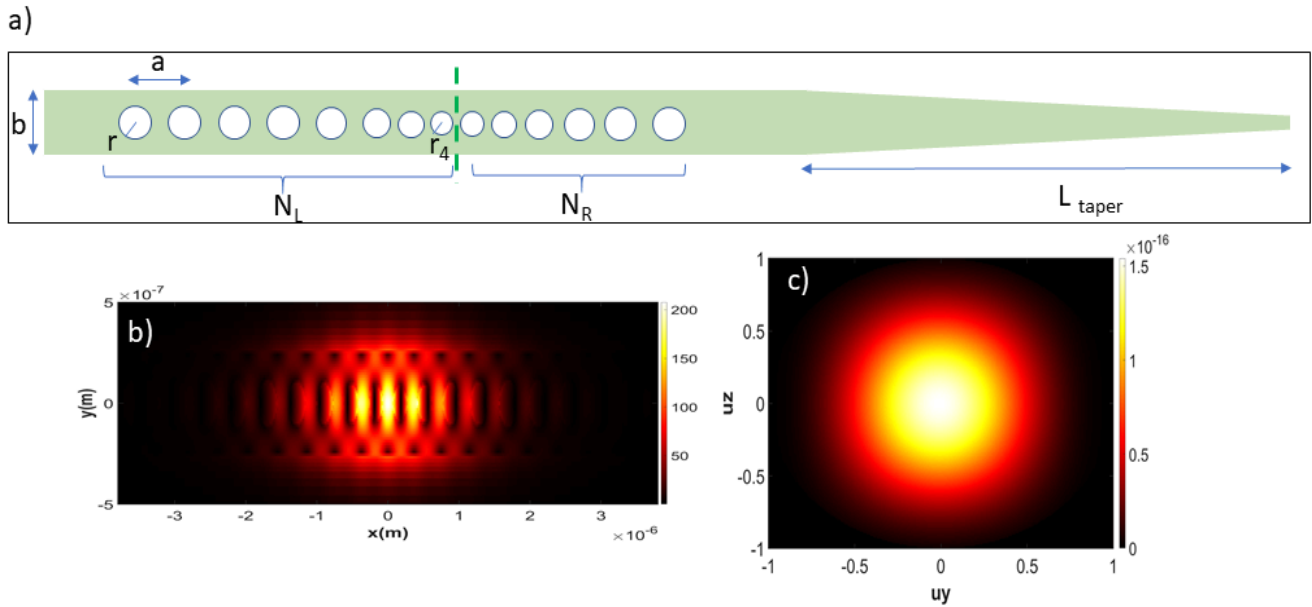

Fig S3: a) Schematic of nanobeam cavity design. b) Electric field profile of the cavity obtained from FDTD simulations. c) Fairfield profile of the mode guided through the adiabatic taper on the right.

We also design bare nanobeam structures with 13 equally spaced holes, created by removing the right-side holes from the cavity design. In this configuration, the left side functions as a full Bragg reflector and guides the light through the adiabatic tapered waveguide on the right to the lensed fiber. This waveguide structure allows us to study the emission properties of Er-doped ceria nanocrystals in the absence of cavity effects for comparison.

# Section 4: Fabrication details

We pattern the nanobeam structures on silicon using electron beam lithography with a beam current of 300 pA. We spin-coat a positive electron-beam resist (ZEP 520) at 2600 rpm for 1 minute to obtain a resist thickness of approximately 400 nm. We bake the sample at 180 degrees for 3 mins. To make the surface conductive and prevent charge accumulation during exposure, we apply a layer of AquaSave.

After exposure, we develop the resist using a cold development process with xylene to transfer the pattern. We then bake the sample at 120 °C for 2 minutes. We etch the silicon layer using fluorine-based reactive ion etching (RIE) and remove the remaining resist using Remover PG.

Finally, we perform a chemical wet etch using a 1.5:5 mixture of hydrofluoric acid(HF) and buffered oxide etchant (BOE) to remove the sacrificial $SiO_2$ layer and suspend the nanobeam structures.

Fig S4 shows an optical image of the suspended nanobeam structure arrays and the zoomed in image is an SEM image of one pad containing 6 nanobeams.

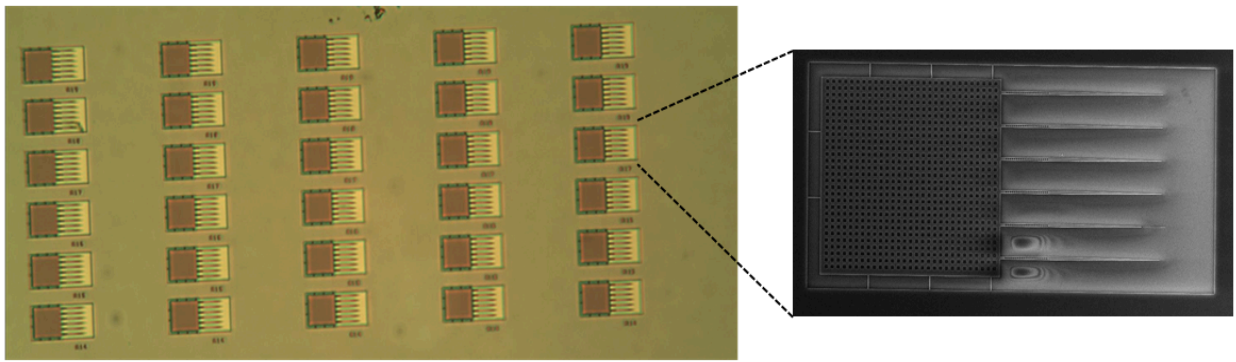

Fig S4 : Optical and SEM image of fabricated nanobeams

We then use a square polydimethylsiloxane (PDMS) microstamp to contact the square pad and rely on van der Waals forces to lift the pad containing the nanobeams by breaking the supporting bridges of the suspended structure. We then place the picked-up pad on the edge of a silicon carrier wafer, aligning it so that the nanobeams protrude perpendicularly from the edge to enable coupling with a lensed fiber.

## Section 5: Measurement setup

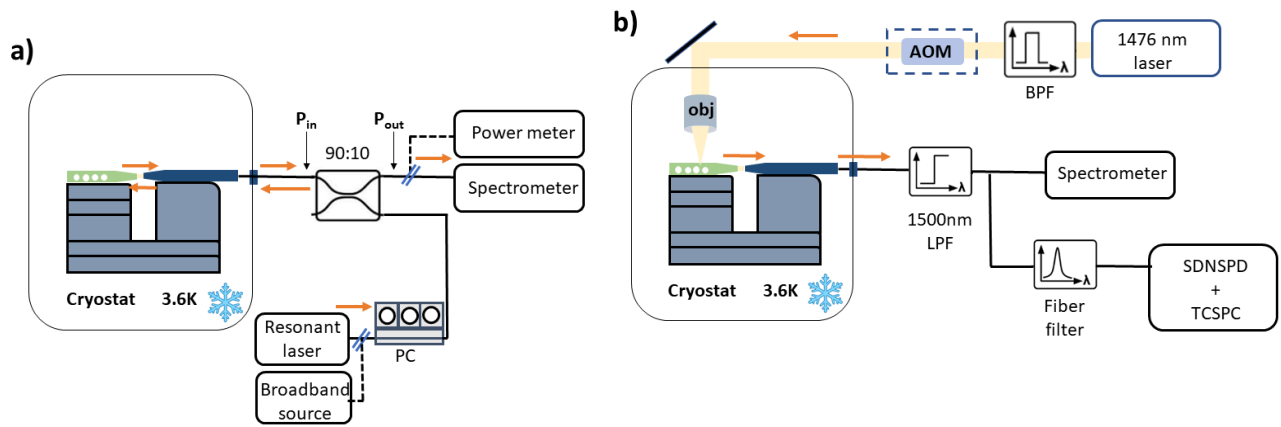

Fig S5: (a) Measurement setup for cavity reflectivity. PC: polarization controller. (b) Measurement setup for emission characterization. SNSPD: superconducting nanowire single photon detectors, TCSPC: time-correlated single photon counter.

Fig S5a shows the reflectivity setup used to obtain the cavity resonance mode as shown in Fig 2d in the main manuscript. We first send in a broadband source along the lensed fiber and measure the reflected light using a 90:10 fiber coupler and a charge coupled device and spectrometer. As our cavity resonance mode is very narrow(high Q), the spectrometer resolution of 0.042nm is not enough and we use a resonant laser scan using a Santec TSL-570 tunable laser across the cavity to characterize the cavity mode (Fig2d in main manuscript).

Figure S5b illustrates the setup for photoluminescence and time-resolved lifetime measurements. The sample is illuminated from above through a microscope objective using a 1476nm diode laser. The emitted light is coupled into a lensed fiber and, after appropriate laser-rejection filtering, directed to either single-photon detectors or a spectrometer for analysis. For lifetime measurements, the excitation is gated with an acousto-optic modulator (AOM) to generate pulses of the desired duration ( 5-ms pulses at a 20% duty cycle).

# Section 6: Coupling efficiency calculation

Assuming that the efficiency of coupling-in and coupling-out of light in the nanobeam is same, we perform reflectivity measurement to estimate the coupling efficiency($\eta$) of the nanobeams to the lensed fiber. We use a 1500 nm narrowband laser and measure the power values using a powermeter. We measure the input ($P_{in}$) and out power ($P_{out}$) at locations shown in Fig S5(a). The off resonance reflectivity of the nanobeams is given by $R = P_{out}/P_{in} = \eta^2 T_{fc}$ where $T_{fc}$ is the actual transmission of 90:10 fiber coupler which is 85% . Assuming that the photonic crystal mirror on left is a perfect Bragg reflector, we get $\eta = 45\% - 60\%$ for different beams. As we dropcast solution of erbium doped nanocrystals on the nanobeams, we attribute this variation in coupling efficiency to the bend introduced in the beams because of solution deposition.

# Section 7: Nanobeam cavity mode shift at 3.6 K compared to room T

Photonic crystal cavities often exhibit a shift in resonance wavelength at cryogenic temperatures compared to room temperature. This shift arises from temperature-dependent changes in the refractive index of the substrate or thermally induced variations in feature dimensions. Accurately measuring this shift is crucial, as it impacts the target resonance wavelength during fabrication.

In our silicon nanobeam cavity design, we typically observe a ~12 nm blueshift in the cavity mode at 3.6 K relative to ambient temperature. However, in our experiments, an additional dried layer of solution on the nanobeams causes the blueshift to vary between 12 nm and 20 nm. This variability complicates efficient coupling of the nanobeam cavities to erbium emission. Fig S7 shows a representative cavity reflectivity spectrum measured using a broadband source. The data illustrate an 18 nm blueshift in the cavity mode at cryogenic temperatures (blue points) compared to the room-temperature mode (orange points).

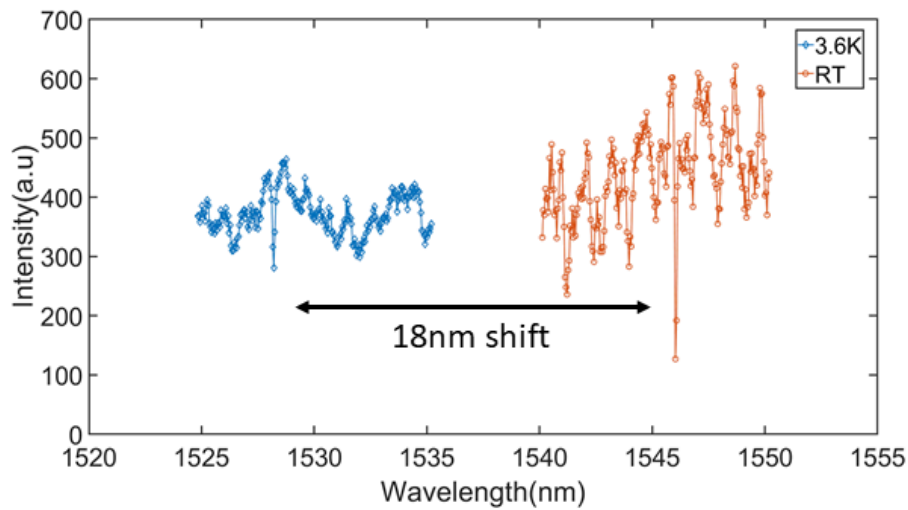

Fig S7: Wavelength shift of nanobeam cavity mode at 3.6K compared to room temperature

## Section 8: Concentration uniformity verification

Since drop-casting is an inherently uncontrolled process, we ensure that approximately equal amounts of solution are deposited on the different beams used for brightness comparison. To uniformly excite all erbium ions throughout the beams—including those in the tapered waveguide—we use a 1476 nm pump from the side through a lensed fiber( Fig S5a). Emission is then collected through the same lensed fiber after filtering out the pump laser. Fig S8 shows the measured emission from a mirror waveguide (green) and a nanobeam cavity (orange). Excluding the cavity-enhanced peak, the background emission within the region marked by the red dashed rectangle shows comparable intensity in both cases, indicating similar emitter concentrations in the two beams.

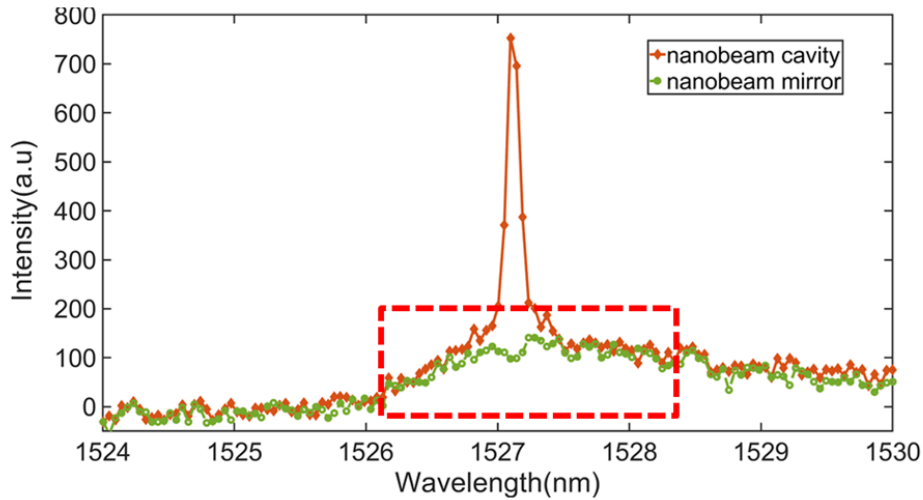

Fig S8: Emission spectra of all erbiums in the entire nanobeam to determine concentration uniformity in two beams.

## Section 9 : Lifetime of annealed nanocrystals drop-casted in nanobeam mirror waveguide

The ensemble-averaged lifetime of erbium emission measured in the nanobeam waveguide is approximately 50 μs—significantly shorter than the millisecond-scale radiative lifetimes reported in other hosts, such as 3 ms in epitaxial $CeO_2$. We attribute this reduced lifetime to dominant non-radiative decay via defect-related quenching pathways. To investigate this, we annealed the sample at 500 °C for 1 hour in a box furnace. Annealing is known to effectively repair strain-induced damage in crystals caused during the doping process.

We performed time-resolved photoluminescence measurements to estimate the decay time of annealed erbium-doped $CeO_2$ in the nanobeam waveguide. Figure S8 shows the decay curve, with the black solid line representing a stretched exponential fit. From the fit, we extract an average decay time of $T_{avg}$ = 320 µs with a stretch parameter β=0.51. This longer lifetime approaches the intrinsic radiative lifetime of erbium, suggesting that annealing effectively reduced non-radiative decay channels.

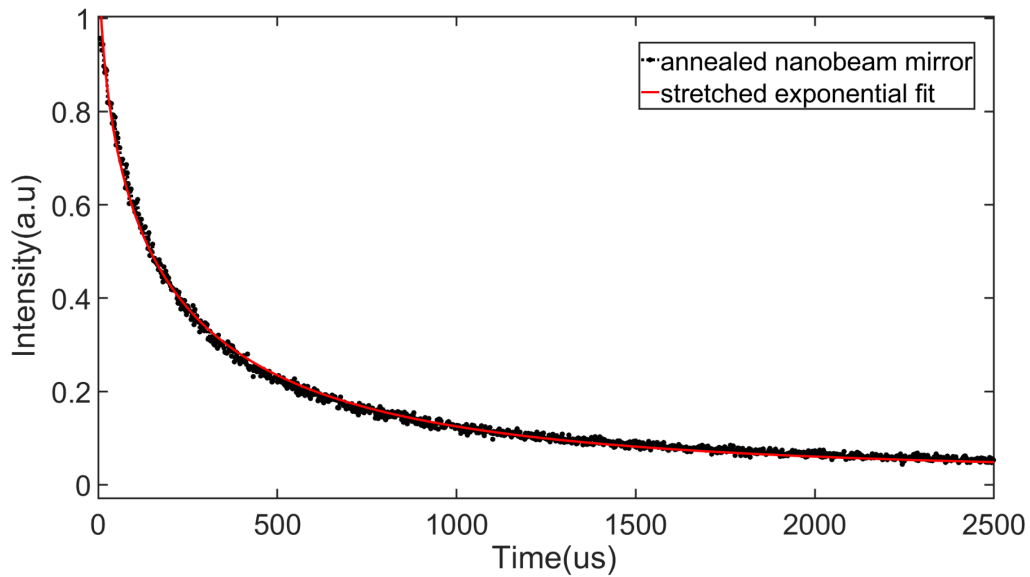

Fig S9: Time resolved lifetime measurement of Er in a nanobeam mirror waveguide after annealing. The red solid line is stretched exponential fit to the data.

## Section 10: Purcell factor Estimation

To get an estimate of the Purcell factor, we need to factor in both the radiative and non-radiative contribution of the measured decay. We follow the analysis as performed in Kim et al.[6] We know that cavities only enhance the radiative part of emission due to Purcell effect. The emission decay rate of erbium ( $\gamma_0$ ) without the presence of any cavity is given by -

$$\gamma_0 = \gamma_r + \gamma_{nr}, \tag{1}$$

where $\gamma_r$ is the radiative rate and $\gamma_{nr}$ is the non radiative decay rate.
Additionally the radiative rate ($\gamma_r$) can be expressed as -

$$\gamma_r = \gamma_z + \gamma_{SB} = \varepsilon_r \gamma_0, \tag{2}$$

and

$$\gamma_z = F_{DW} \gamma_r = F_{DW} \varepsilon_r \gamma_0 \tag{3}$$

where $\gamma_z$ is the decay rate into zero phonon line (ZPL), $\gamma_{SB}$ is the decay rate into phonon sideband, $\varepsilon_r$ is the quantum efficiency of the emitter and $F_{DW}$ is the Debye-Waller factor.

When erbium is in a nanophotonic cavity with the cavity mode at frequency $\omega$ and ZPL emission at frequency $\omega_0$, the measured cavity decay rate is given by :

$$\gamma_{cav}(\omega) = F_P \frac{(\Delta\omega)^2}{4(\omega-\omega_0)^2+(\Delta\omega)^2} \gamma_z + \gamma_{SB} + \gamma_{nr,cav} \tag{4}$$

where $F_P$ is the Purcell factor when emission is on-resonance with cavity mode ($\omega = \omega_0$), $\Delta\omega$ is the cavity linewidth and $\gamma_{nr,cav}$ is the non-radiative decay rate in a cavity.

We measured the decay rate in a mirror waveguide and in a cavity while detuning the measurement window far from the cavity resonance using a spectral filter. In both cases, we observed the same lifetime, indicating that the non-radiative decay rate remains unchanged, i.e., $\gamma_{nr} = \gamma_{nr,cav}$.

So, the off-resonance decay rate ($\gamma_{off}$) when $\omega$ and $\omega_0$ are far apart (first term in equation 4 vanishes) is given by - $\gamma_{off} = \gamma_{SB} + \gamma_{nr}$. The on-resonance decay rate ($\gamma_{on}$) when $\omega = \omega_0$ is given by - $\gamma_{on} = F_P \gamma_z + \gamma_{SB} + \gamma_{nr}$.

The equations can be simplified to get -

$$\gamma_{on} = F_P \gamma_z + \gamma_{off} \tag{5}$$

Or,

$$F_P = \frac{\gamma_{on}-\gamma_{off}}{\gamma_z} = \frac{\gamma_{on}-\gamma_{off}}{\gamma_0}\frac{1}{F_{DW}}\frac{1}{\varepsilon_{rad}} = \tau_0\left(\frac{1}{\tau_{on}}-\frac{1}{\tau_{off}}\right)\frac{1}{F_{DW}}\frac{1}{\varepsilon_{rad}}$$

where $\tau_{on} = 1/\gamma_{on}$ ($\tau_{off} = 1/\gamma_{off}$) is the decay time when the emitter and cavity is on-resonance (off-resonance), and $\tau_0 = 1/\gamma_0$ is the decay time of the erbium ion without any cavity.

In our case, we have $\tau_{on} = 24.72\ \mu s$, $\tau_{off} = \tau_0 = 320\ \mu s$. So,

$$F_p > 11.9 * \frac{1}{F_{DW}}\frac{1}{\varepsilon_{rad}}$$

To calculate the bound on the Purcell factor, we use the post-annealing lifetime, since it is closer to erbium's intrinsic radiative lifetime. This is appropriate because the cavity only modifies the radiative decay rate. So, the lower bound on Purcell factor ($F_P$) *is* ~12.

### References:

1. Wang, D. *et al.* Synthesis and oxygen storage capacity of two-dimensional Ceria nanocrystals. *Angew. Chem. Weinheim Bergstr. Ger.* **123**, 4470–4473 (2011).

2. Ravel, B. & Newville, M. ATHENA, ARTEMIS, HEPHAESTUS: data analysis for X-ray absorption spectroscopy using IFEFFIT. *J. Synchrotron Radiat.* **12**, 537–541 (2005).

3. None Available. Materials data on CeO2 by materials project. LBNL Materials Project; Lawrence Berkeley National Laboratory (LBNL), Berkeley, CA (United States)

https://doi.org/10.17188/1195334 (2020).

4. Ravel, B. *Artemis-Exended Examples: Dopants.15.10. Handling Dopants - Artemis 0.9*.

5. Malinovskii, Y. A. Refined Crystal Structure of Er2O3. *Soviet Physics. Crystallography* **36**, 882–883 (1991).

6. Kim, K.-Y. *et al.* Bright Purcell-enhanced single photon emission from a silicon G center. *Nano Lett.* **25**, 4347–4352 (2025).